\documentclass[12pt]{article}
\long\def\commabs #1\commabsend{#1}
\long\def\commful #1\commfulend{}
\commabs
\commabsend

\usepackage{color, array, colortbl}
\usepackage{graphicx}
\usepackage{subfigure}
\usepackage{amsfonts}
\usepackage{amsmath}
\usepackage{amssymb}
\usepackage{enumitem}

\newcommand{\cancel}[1]{}
\usepackage{hyperref}
\usepackage{wrapfig}

\commabs
\newtheorem{theorem}{Theorem}[section]

\newtheorem{corollary}[theorem]{Corollary}
\commabsend
\newtheorem{observation}[theorem]{Observation}

\commabs
\newcommand{\qed}{\hfill $\Box$ \medbreak}
\newenvironment{proof}{\noindent{\bf Proof.}}{\qed}
\commabsend

\def\loc{x}

\def\Prb{\mathbb{P}}
\def\Expct{\mathbb{E}}

\newcommand{\LF}{\mbox{\small \sc lf}}
\newcommand{\PF}{\mbox{\small \sc pf}}

\begin{document}


\title{Fault-Tolerant Hotelling Games}

\commful
\author{Chen Avin\inst{1} \and Avi Cohen \inst{2} \and Zvi Lotker\inst{1} \and David Peleg\inst{2}}
\authorrunning{Avin, Cohen, Lotker \& Peleg}   
%
\tocauthor{Chen Avin, Avi Cohen, Zvi Lotker, David Peleg}%
\institute{Ben Gurion University of the Negev, Beer Sheva, Israel\\
\email{avin@cse.bgu.ac.il,zvilo@bgu.ac.il}\\
       \and
Weizmann Institute of Science, Rehovot, Israel\\
\email{avi.cohen@weizmann.ac.il,david.peleg@weizmann.ac.il}
 }
\commfulend
\commabs
\author{
Chen Avin\thanks{Ben Gurion University of the Negev, Beer Sheva, Israel.
avin@cse.bgu.ac.il,~ zvilo@bgu.ac.il}
\and
Avi Cohen\thanks{Weizmann Institute of Science, Rehovot, Israel.
avi.cohen@weizmann.ac.il,~ david.peleg@weizmann.ac.il}
\and
Zvi Lotker$^*$
\and
David Peleg$^{\dag}$
}
\commabsend

\maketitle

\begin{abstract}

The $n$-player Hotelling game calls for each player to choose a point
on the line segment, so as to maximize the size of his Voronoi cell.
This paper studies fault-tolerant versions of the Hotelling game.
Two fault models are
\commful
studied. The first
\commfulend
\commabs
studied: line faults and player faults. The first model
\commabsend
assumes that the environment is prone to failure: with some probability, a disconnection occurs at a random point on the line, splitting it into two separate segments and modifying each player's Voronoi cell accordingly. A complete characterization of the Nash equilibria of this variant is provided for every $n$. Additionally, a one to one correspondence is shown between equilibria of this variant and of the Hotelling game with no faults.
The second fault model assumes the players are prone to failure: each player is removed from the game with \mbox{i.i.d.} probability, changing the payoffs of the remaining players accordingly. It is shown that for $n \geq 3$ this variant of the game has no Nash equilibria.

\commful
\commfulend
\end{abstract}

\thispagestyle{empty}

\section{Introduction}

\commabs
\paragraph{Background.}
The Hotelling game,
originated in Hotelling's seminal work in 1929 \cite{hotelling1929stability}, modeled the competition of two servers on a linear market (e.g., two ice-cream vendors on a beach strip) as follows. Two servers choose a location on the line segment $[0,1]$, and the payoff of each server is equal to the length of the segment of points closer to it than the the other server (a.k.a. its Voronoi cell). Hotelling showed that if both servers locate themselves at the center of the line, then a Nash equilibrium would be reached, i.e., a situation in which neither server would rather relocate unilaterally. He next showed that three servers competing on the line reach no equilibrium state - in every configuration of servers there will be one server who can increase its profit by moving.

\begin{figure}[ht]
  \begin{center}
    \includegraphics[width=0.5\textwidth]{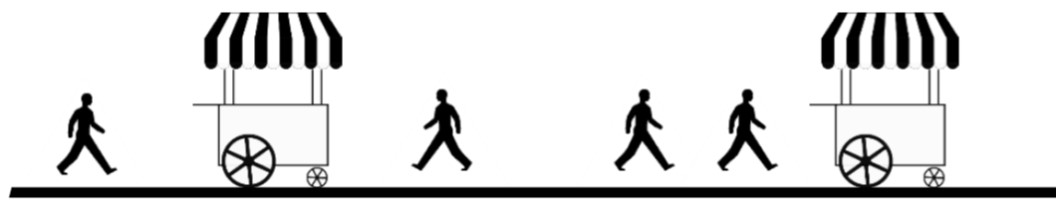}
\caption{Ice cream vendors on a beach strip.}
\label{fig:ice-cream}
  \end{center}
\end{figure}
\commabsend
\commful
\paragraph{Background.}
The Hotelling game,
originated in Hotelling's seminal work in 1929 \cite{hotelling1929stability}, modeled the competition of two servers on a linear market (e.g., two ice-cream vendors on a beach strip) as follows. Two servers choose a location on the line segment $[0,1]$, and the payoff of each server is equal to the length of the segment of points closer to it than the the other server (a.k.a. its Voronoi cell). Hotelling showed that if both servers locate themselves at the center of the line,

\begin{wrapfigure}{R}{0.4\textwidth}
\begin{center}
\includegraphics[width=0.4\textwidth]{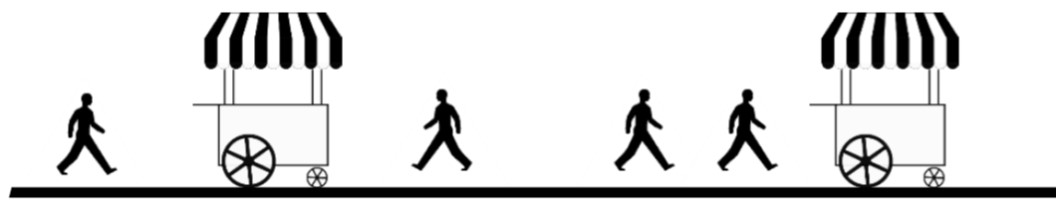}
\end{center}
\vspace{-15pt}
\end{wrapfigure}
\noindent
then a Nash equilibrium would be reached, i.e., a situation in which neither server would rather relocate unilaterally. He next showed that three servers competing on the line reach no equilibrium state - in every configuration of servers there will be one server who can increase its profit by moving.
\commfulend

This initial idea sparked decades of research. Notably, in 1975, Eaton and Lipsey \cite{eaton1975principle} completely characterized all Nash equilibria of the $n$ server game, for every $n$. Over the years numerous variations were made to each component of the game, including the number of players, the pricing policy, the behavior of clients, and the geometry of the market. Eiselt, Laporte and Thisse~\cite{eiselt1993competitive} provide an annotated bibliography categorized by these features (for a more recent survey see Eiselt et al.~\cite{eiselt2011equilibria}).

\paragraph{Motivation.}
In this paper, we consider the Hotelling game in a failure prone setting, and explore how the Hotelling game behaves differently in a world where faults may occur either in the environment, or in the servers themselves. In day to day life, uncertainly must be accounted for. For instance, one of the players may fail to open their store due to illness or vacation, or the road might be blocked due to infrastructure work or safety issues, denying clients access to their preferred vendor. While it is uncertain whether such an event occurs on a given day or not, it is certain to happen one day. It therefore stands to reason that failures would be accounted for in player strategies and payoffs.
Indeed, fault-tolerant problems constitute a fertile area of research in Computer Science. Yet, to the best of our knowledge, this paper is the first to consider fault-tolerance aspects of the Hotelling game.

\paragraph{Contributions.}
We analyzed two types of failure models. In the first variant, we consider a failure prone \emph{environment}, wherein it is possible that the line would be blocked at a random point, denying the passage of clients through it. We refer to this variant of the game with $n$ players as the {\em Line Failure Hotelling game} and denote it by $H_{\LF}(n)$. We characterize all existing equilibria of $H_{\LF}(n)$ (Theorem~\ref{thm:disconnect}). Moreover, we show that each Nash equilibrium of $H_{\LF}(n)$ corresponds to a Nash equilibrium of the non-failure Hotelling game $H(n)$ (Theorem~\ref{thm:equiv}).

The second variant
we consider assumes a reliable environment, but failure prone players. Each player (independently) has some probability of being removed from the game. This version of the game (with $n$ players) is referred to as the {\em Player Failure Hotelling game}, denoted $H_{\PF}(n)$.
We show that
\commful
for $n\ge 3$ players,
\commfulend
\commabs
if there are at least three players, then
\commabsend
the game admits no Nash equilibrium (Theorem~\ref{thm:crash}).

\paragraph{Related Work.}
Fault tolerant facility location problems have been studied extensively from an optimization perspective in the operations research and computer science communities \cite{chechik2014robust,chechik2015fault,khuller2000fault,snyder2016or,sviridenko2002improved,swamy2008fault}. However, relatively little work has been done from a game theoretic approach. Two recent papers do consider fault tolerant location games. Wang and Ouyang studied a failure prone competitive location game in a two dimensional environment~\cite{wang2013continuum}. However, they studied a variant in which two players position several facilities each, and did not extend the model to a greater number of players. Zhang et al. considered a discrete competitive location model in which there are finitely many clients and finitely many potential facility locations~\cite{zhang2016competitive}. Their model, while similar to our model in theme, bears little resemblance to Hotelling's original game.

In 1987, De Palma et al.~\cite{de1987existence} introduced randomness into the Hotelling game, though in a manner different than we do. Rather than uncertainty in the reliability of the environment or the players, their paper studied uncertainty in client behaviors. That is, clients normally shop at the closest vendor, but with some probability, due to unquantifiable factors of personal taste, they skip a seller and travel further to the next one. This paper has a similar approach to our paper, but considers a different problem than ours.

\commful
Other types
\commfulend
\commabs
Aside for location games, other types
\commabsend
of fault tolerant games have recently been studied. For example,
Gradwohl and Reingold \cite{GR14} studied the immunity and tolerance of games with many players.
Immunity means that faults have a small affect on the utility of non-faulty players. Tolerance means that optimal strategies remain optimal when faults are introduced to the game (even though the utilities may be different from the case without faults). The authors show that the \emph{games themselves} are resilient to faults and quantify the strength of their resilience. We, on the other hand, consider a game that is very sensitive to faults and ask how would the players adapt their strategies in a given fault model.

\paragraph{Organization.}
The paper is organized as follows. Section~\ref{sec:model} presents the Hotelling game formally along with known results. Sections~\ref{sec:disconnect} and
\ref{sec:crash} analyze the Hotelling game with line disconnections and server crashes, respectively.
Section~\ref{sec:conclusion} concludes the paper and offers future directions of research.

\section{Model and Preliminaries} \label{sec:model}

\paragraph{The Game.}
The Hotelling game on the line segment $[a,b]$, denoted as $H(n,[a,b])$, involves $n$ servers, $s_i$ for $i=1,\ldots,n$, who place themselves at different points $\loc_i$ on the segment.
Each point on the line also represents a client, who will be served by the nearest server.
The \emph{market} of each server is the line segment containing the clients that
will be served by it (this segment is also known as the server's \emph{Voronoi cell}).
The payoff of each server $s_i$, denoted $p(s_i)$, is the length of his market.
Servers strive to maximize their payoff.

We assume clients are uniformly distributed over the line. We also assume that no more than one
server can occupy a given location; the minimal distance between two servers is some arbitrarily small $\varepsilon>0$.%
\footnote{Setting a minimal distance between servers is common practice in the Hotelling game literature. This prevents servers from infinitely moving closer and closer to each other to slightly improve their payoff. However, the servers choose their location simultaneously and thus two servers might inadvertently choose the same location. We assume that in this case players make small corrections until they meet the constraint. Alternatively, we could say that two servers can be located at the same point and split the payoff in half. However, this leads to unnatural equilibria and thus makes the analysis more complicated. For example, in the two-server game, locating both servers at the same point, anywhere on the line, yields an equilibrium.}
When two servers are separated by a distance of   $\varepsilon$ they are said to be \emph{paired}. We say two servers $s_i$ and $s_j$ are \emph{paired at location} $\loc$ if $\loc_i=\loc-\varepsilon/2$ and $\loc_j=\loc+\varepsilon/2$;
with a slight abuse of notation, we hereafter denote this as $\loc_i=\loc_j=\loc$. Conversely, a server is \emph{isolated} if it is not paired to another server.

\begin{figure}
\includegraphics[width=\textwidth]{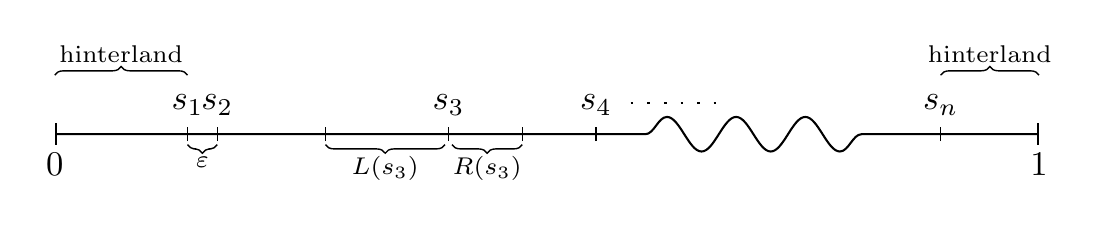}
\caption{A possible configuration of the game. Servers $s_1$ and $s_n$ are the peripheral servers, $s_1$ and $s_2$ are paired, and the half-markets of $s_3$ are marked in the figure.}
\label{fig:Hgame-defs}
\end{figure}

Each server $s_i$ divides its market into two sides, referred to as \emph{half-markets}. We denote by $L(s_i)$ and $R(s_i)$ the lengths of $s_i$'s left and right half-markets respectively. Therefore, the payoff of $s_i$ is
\commful
$p(s_i)=L(s_i)+R(s_i)$.
\commfulend
\commabs
\[  p(s_i)=L(s_i)+R(s_i). \]
\commabsend
Two servers are said to be \emph{neighbors} if no server is located between them. A server that has neighbors on both sides is called an \emph{interior server}. A server that has one neighbor is called a \emph{peripheral server}. That is, the two peripheral servers are the server closest to 0 and the server closest to 1. (See Fig. \ref{fig:Hgame-defs}.)

By definition, the market of an interior server extends half the distance to its two neighbors. The length of an interior server's market is thus half the distance between its neighbors, wherever the server is located between those neighbors. The line segment between the boundary and the corresponding peripheral server is called a \emph{hinterland}. The market of a peripheral server includes its hinterland in one direction, and extends half way to its neighbor in the other direction.

The game is in a \emph{Nash equilibrium} if no server can increase its payoff by moving to a location other than its present location.

\begin{figure}[h]
\includegraphics[width=\textwidth]{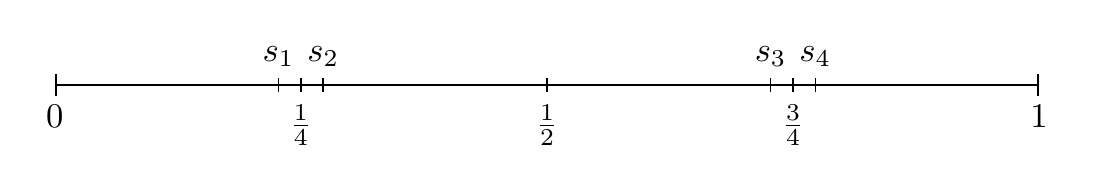}
\caption{Equilibrium with 4 servers.}
\label{fig:Hgame}
\end{figure}

\commabs
As an example, consider the basic Hotelling game which is played on the line $[0,1]$, and involves four servers, $s_i$ for $i=1,2,3,4$, who need to place themselves at different points on the line, $\loc_i$ for $i=1,2,3,4$.
Consider a solution in which servers $s_1$ and $s_2$ place themselves at the point $1/4$, with $s_1$ to the left of $s_2$, and servers $s_3$ and $s_4$ place themselves at the point $3/4$, with $s_3$ to the left of $s_4$. (See Figure \ref{fig:Hgame}.)
It is easy to verify that this solution of the game,
whose payoff is $1/4$ per server, is a Nash equilibrium
(i.e., none of the servers is motivated to change its location).
\commabsend

\paragraph{Known Results.}
Eaton and Lipsey (1975)~\cite{eaton1975principle} proved that the following are necessary and sufficient conditions for an equilibrium:
\begin{enumerate}[label=(EL\arabic*)]
  \item Each peripheral server is paired. \label{p1}
  \item No server's whole market is smaller than any other server's half-market. \label{p2}
\end{enumerate}

The proof these conditions are sufficient is a bit involved, but it is easy to see why they are necessary. If \ref{p1} does not hold, a peripheral server would increase its profit by locating closer to its neighbor. \ref{p2} follows from the fact that any server can obtain a market equal to any other server's half-market by pairing with it.

By applying these equilibrium conditions, Eaton and Lipsey determined the equilibria of the game depending on the number of
\commful
servers, as
\commfulend
\commabs
servers. We present their results in what
\commabsend
follows.

\noindent \emph{One Server.} The payoff of a single server $s_1$ is $p(s_1)=1$,
wherever it is located.

\noindent \emph{Two Servers.}
As shown by Hotelling,
there is a unique equilibrium, where the two servers
are paired at $\loc_1=\loc_2=1/2$, with payoffs $p(s_1)=p(s_2)=1/2$.

\noindent \emph{Three Servers.} No equilibrium exists. It follows from the fact that \ref{p1} and \ref{p2} contradict one another in this case.

\noindent \emph{Four Servers.} There is a unique equilibrium,
with equal payoffs of $1/4$,
where two servers are paired at $\loc_1=\loc_2=1/4$, and the other two
at $\loc_3=\loc_4=3/4$.

\noindent \emph{Five Servers.} There is a unique equilibrium, where two servers are paired at $\loc_1=\loc_2=1/6$, two others are paired at $\loc_4=\loc_5=5/6$, and an isolated server is located at $\loc_3=1/2$. Note that here, the payoffs are not uniform: $s_3$ has payoff $p(s_3)=1/3$, while all other servers have payoff $1/6$.

\noindent \emph{More than five Servers.} There exist infinitely many equilibria, characterized as follows: peripheral servers are paired with their neighbors and have identical hinterlands. Each peripheral pair
is separated from the closest server by a distance twice as long as the hinterland. The interior servers are paired or isolated.

As an example, consider the game $H(6,[0,1])$.
Let the length of the hinterland be $x$, and
without loss of generality let the servers be ordered such that $ \loc_1 < \loc_2 < \ldots < \loc_6 $. From the above characterization of equilibria, it follows that in every equilibrium $s_1$ and $s_2$ are paired at $\loc_1=\loc_2=x$, $s_3$ is located at $\loc_3=3x$, $s_4$ is located at $\loc_4=1-3x$ and $s_5$ and $s_6$ are paired at $\loc_5=\loc_6=1-x$. The distance between $s_3$ and $s_4$ is at least $\varepsilon$ and at most $2x$ (by condition \ref{p2}).

\begin{figure}
\includegraphics[width=\textwidth]{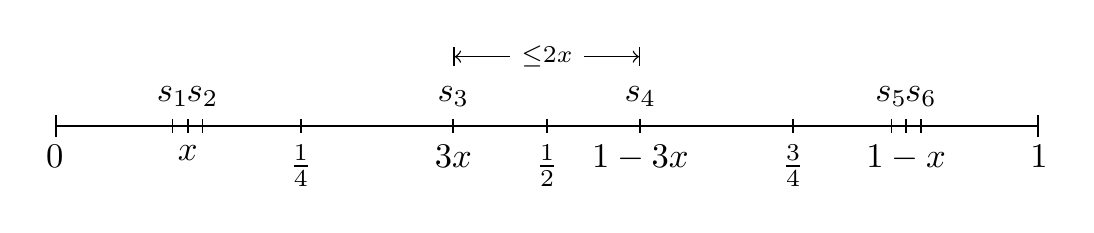}
\caption{Equilibrium with 6 servers.}
\label{fig:Hgame-6p}
\end{figure}

\noindent
That is, the above described configuration is an equilibrium for every $x$ such that
\commful
$\varepsilon \leq (1-3x)-3x \leq 2x$,
\commfulend
\commabs
$$\varepsilon \leq (1-3x)-3x \leq 2x~,$$
\commabsend
or rather $1/8 \leq x < 1/6$.
(See Figure \ref{fig:Hgame-6p}.)

\vspace{-12pt}

\section{Hotelling Game on the Line with Link Failures} \label{sec:disconnect}

Let us now consider the game with
environmental failures. For concreteness, let us assume that the only possible failure is a disconnection of the line at point $f$, chosen uniformly from $[0,1]$, which severs the line into two separate markets, and forces some of the clients (specifically, those disconnected from their originally chosen server) to change their server selection. For simplicity, assume at most one disconnection may occur, with constant probability $0<r<1$, at a location $f\in[0,1]$ chosen uniformly at random. We call this game the \emph{Line Failure Hotelling game} and denote it as $H_{\LF}(n,r,[0,1])$. The new payoff function is denoted as $p_{\LF}$ and becomes the expected profit under these assumptions (i.e., the payoff of player $s_i$ is $1-r$ times its payoff in the fault-free case plus $r$ times its expected payoff in case a disconnection occurred at point $0<f<1$).

Let us begin by considering this game with only \emph{one server}. Unlike the basic Hotelling game, in which the location of a single server is inconsequential, in this setting the optimal location of a single server is at the center of the line, $\loc_1=1/2$, as we show next.
Let $\loc_1\in [0,1]$ be the location of the server $s_1$. If no failure occurs, the payoff is 1.
If the line is disconnected at $0<f<\loc_1$, then the payoff is $1-f$.
If the line is disconnected at $\loc_1<f<1$, then the payoff is $f$.
It follows that the payoff of $s_1$ is
$$p_{\LF}(s_1)=\Expct[p(s_1)]=(1-r)\cdot 1+ r\cdot\left[\int_{0}^{\loc_1}(1-f)df+\int_{\loc_1}^{1}f df\right]
= 1-\frac r2+r\loc_1-r\loc_1^2~.$$

That is, $ p_{\LF}(s_1) $ is a function of $\loc_1$ that attains its maximum at $\loc_1=1/2$.
Hence the only equilibrium is when the server is
at the center of the market.

\begin{figure}[h]
\includegraphics[width=\textwidth]{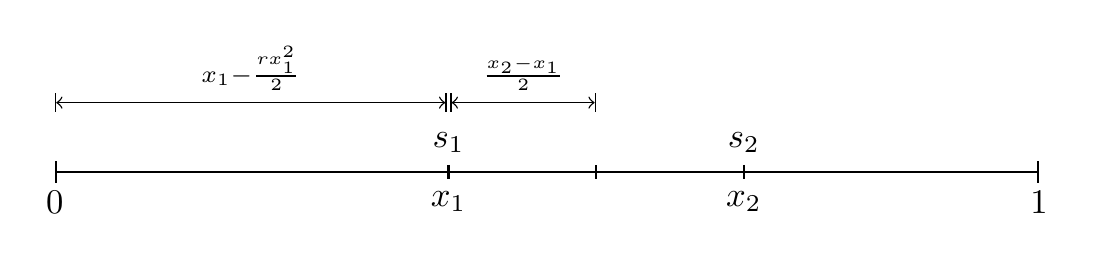}
\caption{Calculating the expected payoff in the two-player game with link failures.}
\label{fig:Hgame-2p}
\end{figure}

We next consider the \emph{two server} variant. Without loss of generality let $\loc_1<\loc_2$, i.e.,
$s_1$ is located to the left of $s_2$. Observe that the payoff of $s_1$ is $p_{\LF}(s_1)=L_{\LF}(s_1)+R_{\LF}(s_1)$, where $L_{\LF}(s_1)$ is the length of its hinterland, and $R_{\LF}(s_1)$ is the length of its half-market on the right.
\commabs
(See Fig. \ref{fig:Hgame-2p}.)
\commabsend

\begin{itemize}
\item
\textbf{Calculating $\mathbf{L_{\LF}(s_1)}$:} if the line is disconnected at location $0<f<\loc_1$,
then the length of $s_1$'s hinterland is $L(s_1)=\loc_1-f$. Otherwise, $L(s_1)=\loc_1$.
The expected length is therefore
$$
    L_{\LF}(s_1) =
 \Expct[L(s_1)] = (1-r)\cdot\loc_1
                +r\cdot\left[\int_{0}^{\loc_1}(\loc_1-f)df+\int_{\loc_1}^{1}\loc_1 df\right]
                =\loc_1-\frac{r\cdot\loc_1^2}2~.
$$
\item
\textbf{Calculating $\mathbf{R_{\LF}(s_1)}$:} if no failure occurs, or if the line is disconnected at location $0<f<\loc_1$ or $\loc_2<f<1$, then the length of the half-market is
$R(s_1)=(\loc_2-\loc_1)/2$. If, on the other hand, an edge is disconnected at $\loc_1<f<\loc_2$, then
$R(s_1)=f-\loc_1$. Therefore,
$$
    R_{\LF}(s_1) =
    \Expct[R(s_1)]=
    (1-r)\cdot\frac{\loc_2-\loc_1}2+r\cdot\frac{\int_{\loc_1}^{\loc_2}(f-\loc_1)df}{\loc_2-\loc_1}=
    \frac{\loc_2-\loc_1}2~.
$$
\end{itemize}

\commabs
\commabsend

It follows that the payoff of $s_1$ is
$$
    p_{\LF}(s_1)=L_{\LF}(s_1)+R_{\LF}(s_1)=
    \loc_1-\frac{r\cdot\loc_1^2}2+\frac{\loc_2-\loc_1}2=\frac{\loc_2+\loc_1}2-\frac{r\cdot\loc_1^2}2~.
$$
This function attains its maximum at $\loc_1=1/(2r)$.
That is, as long as $s_1$ remains on the left of $s_2$, $s_1$ would move to $1/(2r)$.
If $1/2 \leq \loc_2 < 1/(2r)$, then $s_1$'s best response would be pairing with $s_2$.

\begin{figure}[h]
\includegraphics[width=\textwidth]{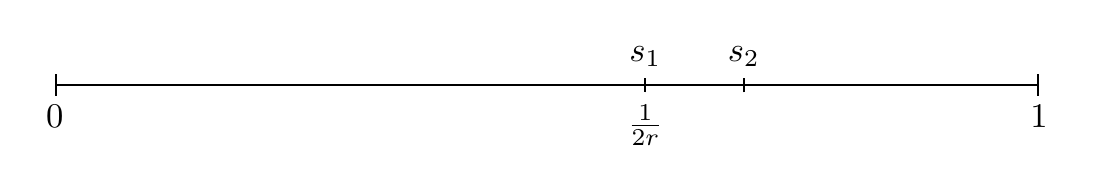}
\caption{$s_1$ is located such that its payoff is maximal, and moving closer to $s_2$ would reduce $s_1$'s payoff. However, $s_2$ would maximize its payoff by moving to the left of $s_1$.}
\label{fig:Hgame-2p-max}
\end{figure}

Note that $1/(2r)>1/2$ for every $0<r<1$ and thus if $\loc_1=1/(2r)$ and
$\loc_2>1/(2r)$, then $s_2$ would prefer to move to the left of $s_1$.
It follows that $\loc_2<1/(2r)$. By symmetry, it also holds that $\loc_1>1-1/(2r)$.
Hence, condition \ref{p1} holds, and as in the basic Hotelling game,
the only equilibrium is when both servers are paired at the center of the line, i.e., $\loc_1=\loc_2=1/2$.
\commabs
(See Fig. \ref{fig:Hgame-2p-max}.)
\commabsend

\commful
Consequently,
\commfulend
\commabs
As a result of the above calculations,
\commabsend
for the general game, with $n\geq 3$ servers, we have the following.

\begin{observation} \label{obs:disconnect}
In the Hotelling game $H_{\LF}(n,r,[0,1])$,
\commabs
on the line with edge disconnections,
\commabsend
the following holds:
\begin{enumerate}
  \item
  A peripheral server located at distance $\loc$ from the boundary has a hinterland with an expected length of
  \commful
  $\loc - r\loc^2/2$.
  \commfulend
  \commabs
  $$\loc - \frac{r\cdot \loc^2}2~.$$
  \commabsend
  \item
    Two neighboring servers at distance $x$
    gain a half-market of expected length $x/2$ each in the direction
    of the other (as in the basic Hotelling game).
  \item
  Each peripheral server would increase his hinterland up to $1/(2r)$. That is, condition
  \ref{p1} holds unless the neighbor of a peripheral server is at a distance of more than
  $1/(2r)$ from the boundary.
\end{enumerate}
\end{observation}

In light of the observation above, if the game is played with \emph{three servers}, then no equilibrium exists. To see why, observe that if the interior server $s_2$ is located between $1-1/(2r)$ and $1/(2r)$, then the peripheral servers $s_1$ and $s_3$ would pair with $s_2$ on both sides leaving it with 0 payoff, and thus $s_2$ would move.
If, on the other hand, $s_2$ is located at $\loc_2>1/(2r)$, then $s_1$ would locate at $1/(2r)$ and $s_3$ would pair with $s_2$. But then $s_3$ could improve its payoff and would thus move.
Due to symmetry, a similar argument holds if we suppose $s_2$ is located at $\loc_2<1-1/(2r)$.


Next, consider the \emph{four server} game $H_{\LF}(4,r,[0,1])$.
Let us consider the strategy profile where the location $0<x<1$ satisfies that
when $s_1$ and $s_2$ are paired at $x$
and $s_3$ and $s_4$ are paired at $1-x$ all four servers receive the same
expected payoff (any server can pair with any other server and get his payoff).
Since $s_1$ and $s_2$ are symmetric to $s_4$ and $s_3$ respectively it suffices to compare the expected payoffs of $s_1$ and $s_2$.

The following table compares the payoffs $p(s_1)$ and $p(s_2)$ according to possible values of $f$, as well as with the failure-free case.

\begin{center}
\begin{tabular}{| c | c | c | c |}
  \hline
  $f$ & $\Prb[f]$ &  $ s_1 $ &  $ s_2 $ \\  \hline
  no failure & $1-r$ & $x$ & $1/2 -x $\\
  $[0,x]$ & $rx$ & $ x/2 $ & $ 1/2 -x $ \\
  $[x,1-x]$ & $r(1-2x)$ & $x$ & $ 1/2 -x $\\
  $[1-x,1]$ & $rx$ & $x$ & $ 1/2 -x $ \\
  \hline
\end{tabular}
\end{center}

\vspace{2ex}

We now solve $\Expct[p(s_1)] = \Expct[p(s_2)] $ for $x$, namely,
\[
(1-r)\cdot x + rx \cdot \frac x2 + rx \cdot x + r(1-2x) \cdot x = \frac12 - x~.
\]
This yields
\commful
$rx^2 - 4x + 1 = 0$,
\commfulend
\commabs
\[rx^2 - 4x + 1 = 0 \text{,}\]
\commabsend
and recalling that $0<x<1$ we obtain
\commful
$x=(2-\sqrt{4-r})/r$.
\commfulend
\commabs
\[x=\frac{2-\sqrt{4-r}}{r} \text{.}\]
\commabsend

\begin{figure}[h]
\includegraphics[width=\textwidth]{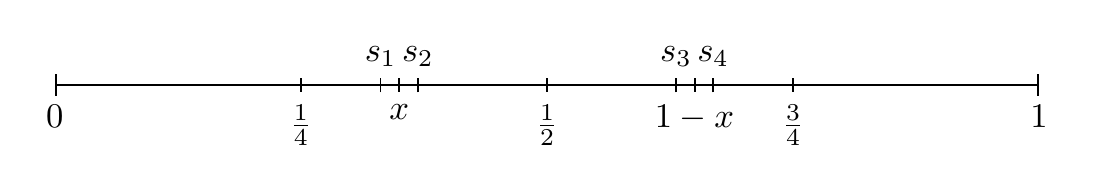}
\caption{Equilibrium with 4 servers ($x=\frac{2-\sqrt{4-r}}{r}$).}
\label{fig:Hgame-4p-Disconnections}
\end{figure}

By the definition of $x$, this means that for every $ r $,  when
$\loc_1=\loc_2=(2-\sqrt{4-r})/r$ and $\loc_3=\loc_4=1-(2-\sqrt{4-r})/r$%
\commabs
\space (See Fig. \ref{fig:Hgame-4p-Disconnections})%
\commabsend
, each server $s_i$ receives the same expected payoff: $\Expct[p(s_i)]=1/2 - \loc $.
We claim that this is a Nash equilibrium, since no server may increase his payoff by moving to another location. This follows from Observation~\ref{obs:disconnect}.
\commabs
As a motivating exercise, in Appendix~\ref{app:nash4} we explicitly show that no improving move exists.
\commabsend
\commful
The detailed calculation is deferred to the full paper.
\commfulend

\commabs
\commabsend

Moreover, this Nash equilibrium is unique due to the following considerations.
First, by Observation~\ref{obs:disconnect}, the peripheral servers must be paired with their neighbors.
Second, the peripheral servers must have hinterlands of the same length,
otherwise one would take the hinterland of the other.
Third, every two paired servers must have the same expected payoff,
otherwise one would take the half-market of the other.
It follows that the configuration above is the only Nash equilibrium.

\begin{figure}[h]
\includegraphics[width=\textwidth]{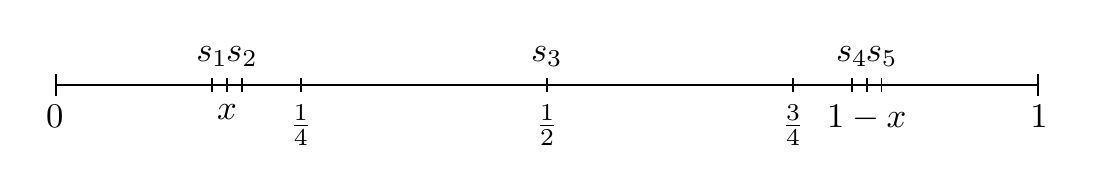}
\caption{Equilibrium with 5 servers ($x=\frac{3-\sqrt{9-4r}}{2r}$).}
\label{fig:Hgame-5p-Disconnections}
\end{figure}

Applying the considerations above to the game with \emph{five servers}
we find that the only equilibrium is when $s_1$ and $s_2$ are paired at $\loc_1=\loc_2=x$, $s_3$ is isolated at $\loc_3=1/2$, and $s_4$ and $s_5$ are paired at $\loc_4=\loc_5=1-x$, such that the distance between $s_3$ and each peripheral pair is exactly twice $x- r \cdot x^2/2$.
\commabs
(See Fig. \ref{fig:Hgame-5p-Disconnections}.)
\commabsend

Hence $x$ satisfies
$1/2 - x = 2 \cdot (x - r x^2/2)$,
which yields
$r x^2 -3x + 1=0$,
and recalling that $0<x<1$, we obtain
\commful
$x=(3-\sqrt{9-4r})/(2r)$.
\commfulend
\commabs
Hence $x$ satisfies
$$\frac12 - x = 2 \cdot \left( x- \frac {r \cdot x^2}2 \right)~,$$
which yields
$$r x^2 -3x + 1=0~,$$
and recalling that $0<x<1$, we obtain
$$x=\frac{3-\sqrt{9-4r}}{2r}~.$$
\commabsend

\commabs
\commabsend

We next consider the general $n$-server game $H_{\LF}(n,r,[0,1])$. By Obs.~\ref{obs:disconnect}, for any given configuration of servers, adding link failures to the game only affects the expected payoff gained from the hinterlands. Namely, a hinterland of length $x$ shrinks by $r x^2/2$ in expectation, while every other half-market retains its length in expectation. This leads to the following theorem. (Hereafter, proofs are deferred to the full paper.)
Let $a=r x_1^2/2$ and $b=1-r x_n^2/2$, hence
$$[a,b] ~=~ \left[\frac{r \cdot x_1^2}2,1-\frac{r \cdot (1-x_n)^2}2\right]~.$$

\begin{theorem}\label{thm:equiv}
Let $0\leq x_1\leq x_2\leq\ldots\leq x_n\leq 1$. The configuration $(x_1,x_2,\ldots,x_n)$ is a Nash equilibrium of the game $H_{\LF}(n,r,[0,1])$ if and only if
it is a Nash equilibrium of the game $H(n,[a,b])$.

\end{theorem}



\commabs

\begin{figure}[h]
\includegraphics[width=\textwidth]{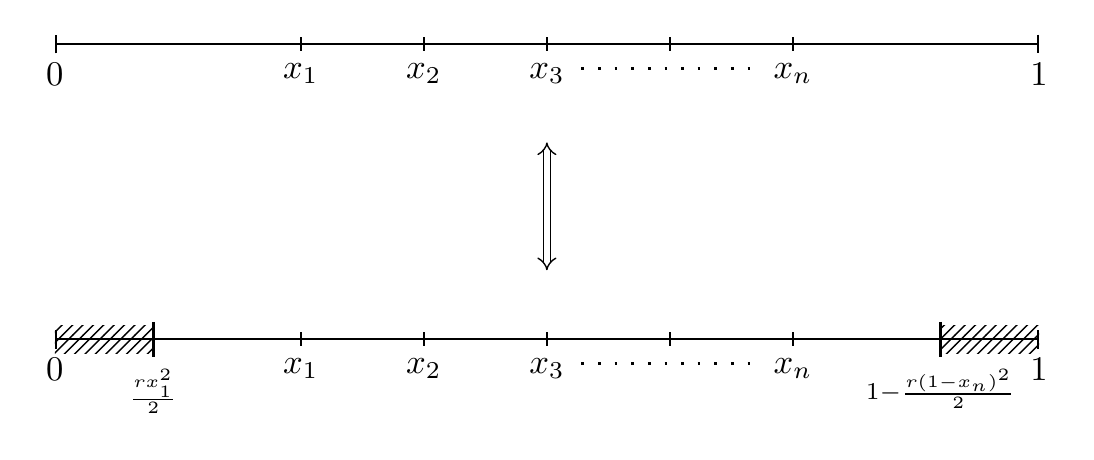}
\caption{Illustration of Theorem~\ref{thm:equiv}. The top line refers to the configuration in the game $H_{\LF}(n,r,[0,1])$ and the bottom line refers to the configuration in the game $H(n,\left[\frac{r \cdot x_1^2}2,1-\frac{r \cdot (1-x_n)^2}2\right])$~. }
\label{fig:Hgame-np-Disconnections}
\end{figure}

\begin{proof}
Suppose that the servers $s_1,s_2,\ldots, s_n$ are located at $x_1,x_2,\ldots,x_n$ respectively.
Note first that both peripheral servers $s_1$ and $s_n$ must be paired with their respective neighbors. In the forward direction of the claim this holds by Observation~\ref{obs:disconnect}; in the inverse direction this follows from condition \ref{p1}.

We start by showing that the payoff of each server given the configuration $(x_1,x_2,\ldots,x_n)$ is the same in both games $H_{\LF}(n,r,[0,1])$ and $H(n,[a,b])$, i.e., for each $s_i$, $p_{\LF}(s_i)=p(s_i)$.
Consider an interior server $s_i$, for $2\leq i\leq n-1$. In the game $H_{\LF}(n,r,[0,1])$, its payoff is
$$
    p_{\LF}(s_i)=\frac{x_{i+1}-x_{i-1}}2~,
$$
by Observation~\ref{obs:disconnect}. In the game $H(n,[a,b])$, its payoff is also
$$
    p(s_i)=\frac{x_{i+1}-x_{i-1}}2~.
$$

Consider the payoff of the peripheral server $s_1$. The servers $s_1$ and $s_2$ are paired, and thus, by Observation~\ref{obs:disconnect}, the payoff of $s_1$ in the game $H_{\LF}(n,r,[0,1])$ is
$$
    p_{\LF}(s_1)=x_1-\frac{r \cdot x_1^2}2~.
$$
In the game $H(n,[a,b])$, the left border of the line is at $a=\frac{r \cdot x_1^2}2$ and thus the payoff of $s_1$ is also
$$
    p(s_1)=x_1-\frac{r \cdot x_1^2}2~.
$$
Similarly, the server $s_n$ gets the same payoff in both games. This proves that, in the given configuration $(x_1,x_2,\ldots,x_n)$, each server gets the same payoff in both games $H_{\LF}(n,r,[0,1])$ and $H(n,[a,b])$.

We next show that for each move available to the server $s_i$, for $1\leq i \leq n$, in one of the two games $H_{\LF}(n,r,[0,1])$ and $H(n,[a,b])$, there exists a corresponding move in the other game for which $s_i$ obtains a greater or equal payoff.

Consider first a move within the interval defined by $s_i$'s present neighbors. If $s_i$ is an interior server then in both games $H_{\LF}(n,r,[0,1])$ and $H(n,[a,b])$ such a move does not affect $s_i$'s payoff, as we have previously shown. If $s_i$ is a peripheral server, since it must be paired with its neighbor, such a move reduces $s_i$'s payoff.

Second, consider a move to an interior interval between two new neighbors $s_j$ and $s_{j+1}$. In both games, the payoff is equal to half the length of the interval, and thus the move yields the same payoff in both games.

Finally, consider a move to either one of the hinterlands. In both games, the best a server $s_i$ could do is pair with the peripheral server, in which case $s_i$ gets the payoff of this peripheral server. But, as we have shown above, the payoff of each peripheral server is the same in both games $H_{\LF}(n,r,[0,1])$ and $H(n,[a,b])$, and thus $s_i$ obtains the same payoff by making this move in both game. This proves that for each move $s_i$ may make in one of the games $H_{\LF}(n,r,[0,1])$ and $H(n,[a,b])$, there exists a move in the other game for which $s_i$ gets greater or equal payoff.

Assume towards contradiction that $(x_1,x_2,\ldots,x_n)$ is a Nash equilibrium in the game $H_{\LF}(n,r,[0,1])$ and but not in the game $H(n,[a,b])$. It follows that there exists a server $s_i$ that can improve his payoff by moving in the game $H(n,[a,b])$. But, as we have shown above, $s_i$ has the same payoff before moving in both games, and by moving $s_i$ can increase its payoff at least as much in the game $H_{\LF}(n,r,[0,1])$ as in the game $H(n,[a,b])$. Therefore, $s_i$ would also move in the game $H_{\LF}(n,r,[0,1])$, contradicting the assumption.
The inverse direction is similar. This concludes the proof that $(x_1,x_2,\ldots,x_n)$ is a Nash equilibrium in the game $H_{\LF}(n,r,[0,1])$ if and only if it is a Nash equilibrium in the game $H(n,[a,b])$.
\end{proof}
\commabsend

Theorem~\ref{thm:equiv}, in conjunction with conditions \ref{p1} and \ref{p2} for an equilibrium of the basic Hotelling game, yield the following conditions for an equilibrium of the Hotelling game with line disconnections.

\begin{corollary}
The following conditions are sufficient and necessary for an equilibrium of the Hotelling game
on the line with line disconnections.
\begin{enumerate}[label=(\arabic*)]
  \item The peripheral servers are paired, and are located at an identical distance from the boundary, $x$. \label{p1d}
  \item No interior server's whole market is smaller than $x- r \cdot x^2/2$. \label{p2d}
  \item No interior server's half market is larger than $x- r \cdot x^2/2$. \label{p3d}
\end{enumerate}
\end{corollary}

\commabs
\begin{proof}
Suppose $(x_1 ,x_2 ,\ldots,x_n )$, for $x_1\leq x_2\leq\ldots\leq x_n$, is an equilibrium of $H_{\LF}(n,r,[0,1])$. Without loss of generality, let the server $s_i$ be located at $x_i$, for $1\leq i\leq n$. By Theorem~\ref{thm:equiv}, it follows that $(x_1 ,x_2 ,\ldots,x_n )$ is an equilibrium of $H(n,[a,b])$, where $a=r x_1^2/2$ and $b=1-r x_n^2/2$ as above. Therefore, conditions \ref{p1} and \ref{p2} must hold for this configuration in $H(n,[a,b])$.

By condition \ref{p1}, the servers $s_1$ and $s_2$ are paired and the servers $s_{n-1}$ and $s_n$ are paired. By condition \ref{p2}, the hinterlands are equal in length, so
$$
    x_1-\frac{r \cdot x_1^2}2=(1-x_n)-\frac{r \cdot (1-x_n)^2}2~,
$$
which yields
$$
    x_1=(1-x_n)~.
$$
This proves condition \ref{p1d} holds.

By condition \ref{p2}, no interior server's whole market is smaller than any other server's half market, including the hinterland. This proves condition \ref{p2d} holds.

By condition \ref{p2}, no interior server's half-market is larger than any other server's half market, including the hinterland. This proves condition \ref{p3d} holds.

Conversely, suppose the configuration $(x_1 ,x_2 ,\ldots,x_n )$ satisfies conditions \ref{p1d},\ref{p2d} and \ref{p3d}, for $x_1\leq x_2\leq\ldots\leq x_n$, and without loss of generality, the server $s_i$ is located at $x_i$, for $1\leq i\leq n$. We show that $(x_1 ,x_2 ,\ldots,x_n )$ is an equilibrium in the game $H(n,[a,b])$.

Given condition \ref{p1d}, condition \ref{p1} follows immediately in the game $H(n,[a,b])$. Furthermore, it holds that
$$
    x_1=(1-x_n)=x~.
$$
By condition \ref{p2d}, the whole market of each interior server $s_i$, for $2\leq i\leq n$ is at least $x- r \cdot x^2/2$, and by condition \ref{p3d} the half market of each interior server $s_j$ is at most $x- r \cdot x^2/2$, for $2\leq j\leq n$. By transitivity, it follows that the whole market of $s_i$ is at least the size of the half-market of $s_j$. This proves that condition \ref{p2} holds.

It follows that  $(x_1 ,x_2 ,\ldots,x_n )$ is an equilibrium of the game $H(n,[a,b])$, and thus, by Theorem~\ref{thm:equiv}, $(x_1 ,x_2 ,\ldots,x_n )$ is an equilibrium of the game $H_{\LF}(n,r,[0,1])$ as well. This concludes the proof of the Corollary.
\end{proof}
\commabsend


To summarize, this section established the following theorem.

\commful
\begin{theorem}\label{thm:disconnect}
In the $n$-server Hotelling game $H_{\LF}(n,r,[0,1])$
  \begin{enumerate}[label=(\roman*)]
    \item
    For $n=1$, a Nash equilibrium exists with $\loc=1/2$.
    \item
    For $n=2$, there exists an equilibrium with $\loc_1=\loc_2=1/2$.
    \item
    For $n=3$, no equilibrium exists.
    \item
    For $n=4$, there exists an equilibrium with
    $\loc_1=\loc_2=(2-\sqrt{4-r})/r$, $\loc_3=\loc_4=1-(2-\sqrt{4-r})/r$.
    \item
    For $n=5$, there exists an equilibrium with
    $\loc_1=\loc_2=(3-\sqrt{9-4r})/(2r)$, $\loc_4=\loc_5=1-(3-\sqrt{9-4r})/(2r)$
    and $\loc_3=1/2$.
    \item
    For $n\ge 6$, there exist infinitely many equilibria,
    characterized as follows:
    peripheral servers are paired with their neighbor and have identical hinterlands of length $x$. Each peripheral pair is separated from the closest server by a distance twice as long as
    $x- r \cdot x^2/2$.
    The interior servers are paired or isolated such that no server's whole market is smaller than
    $x- r \cdot x^2/2$, and no server's half-market is larger than $x- r \cdot x^2/2$.
  \end{enumerate}
\end{theorem}
\commfulend
\commabs
\begin{theorem}\label{thm:disconnect}
In the $n$-server Hotelling game $H_{\LF}(n,r,[0,1])$
with a single line disconnection occurring with probability $0<r<1$ at a location chosen uniformly at random:
  \begin{enumerate}[label=(\roman*)]
    \item
    When there is a single server, a Nash equilibrium exists for which the server is located
    at the center of the line, i.e., $\loc=1/2$.
    \item
    When two servers compete, there exists an equilibrium for which they are paired at the center
    of the line, i.e., $\loc_1=\loc_2=1/2$.
    \item
    If there are three servers, then no equilibrium exists.
    \item
    With four servers, there exists an equilibrium for which the servers are organized in two pairs, at
    $$
        \loc_1=\loc_2=\frac{2-\sqrt{4-r}}{r}~,
        \mbox{\hskip 1cm}
        \loc_3=\loc_4=1-\frac{2-\sqrt{4-r}}{r}~.
    $$
    \item
    If $n=5$, then there exists an equilibrium for which four of the servers are paired at
    $$
        \loc_1=\loc_2=\frac{3-\sqrt{9-4r}}{2r}~,
        \mbox{\hskip 1cm}
        \loc_4=\loc_5=1-\frac{3-\sqrt{9-4r}}{2r}~,
    $$
    and an isolated server is located at $\loc_3=1/2$.
    \item
    When more than five servers occupy the line, there exist infinitely many equilibria,
    characterized as follows:
    peripheral servers are paired with their neighbor and have identical hinterlands of length $x$. Each peripheral pair is separated from the closest server by a distance twice as long as
    $x- r \cdot x^2/2$.
    The interior servers are paired or isolated such that no server's whole market is smaller than
    $x- r \cdot x^2/2$, and no server's half-market is larger than $x- r \cdot x^2/2$.
  \end{enumerate}
\end{theorem}
\commabsend

\section{Hotelling Game on the Line with Server Crashes}\label{sec:crash}

Let us now consider a different failure setting, where the environment is resilient,
but the servers might crash. For concreteness, let us assume that each server
might fail with probability $0<r<1$ independently of the others.
Once a server has failed, the clients who chose this server originally
must change their server selection.
We call this game the \emph{Player Failure Hotelling game} and denote it as $H_{\PF}(n,r,[0,1])$.
The new payoff function is again the expected profit under these assumptions and is denoted as $p_{\PF}$.

\commful
For $n=1$,
\commfulend
\commabs
If a \emph{single server} $s_i$ occupies the line, then
\commabsend
it is clear that the server's expected payoff is $p_{\PF}(s_1)=1-r$
wherever its location is.
\commful
For $n=2$,
\commfulend
\commabs
If there are \emph{two servers}, then
\commabsend
it is easy to see that server crashes have no
\commful
impact;
\commfulend
\commabs
impact on the game. When one server fails, the other inherits the entire line; when both fail, they both get nothing. Hence,
\commabsend
the game is equivalent to the basic Hotelling game and
the only Nash equilibrium is when the servers are paired at the center.

Let us now analyze the case where there are \emph{three servers} on the line.
In every equilibrium, the peripheral servers are
paired with the interior server on both sides, and the interior server is located at $1/2$.
This is because when pairing with the interior server, each peripheral server is closest to its neighbor regardless of which servers crash. Moreover, if they are not located at the center,
then one peripheral server could improve its payoff by taking the hinterland of the other.
Note that, in this state,
the expected payoff of the peripheral server $s_1$ is
$$
    p_{\PF}(s_1)= \Expct[p(s_1)] = (1-r)\left(\frac12+\frac12\cdot r^2\right)~,
$$
since it always gets its hinterland (provided it did not fail), and it gets the remainder of the line only if both other servers have failed.
By a similar case analysis, the expected payoff of the interior server $s_2$ is
$$
    p_{\PF}(s_2) = \Expct[p(s_2)] = (1-r)\left(0+\frac12\cdot r(1-r)\cdot2+1\cdot r^2\right)~,
$$
since (provided it did not fail) it gets nothing if both other servers do not fail, it gets half the line if one of the other servers has failed, and it gets the entire line if both other servers have failed.
Note that $s_2$ can move to take $s_1$'s hinterland,
hence this state would be a Nash equilibrium only if
\commful
$p_{\PF}(s_1)=p_{\PF}(s_2)$,
\commfulend
\commabs
$$p_{\PF}(s_1)=p_{\PF}(s_2)~,$$
\commabsend
i.e.,
$$(1-r)\left(\frac12+\frac12\cdot r^2\right)=(1-r)\left(0+\frac12\cdot r(1-r)\cdot2+1\cdot r^2\right)~,$$
which yields
\commful
$r^2-2r+1=0$,
\commfulend
\commabs
$$r^2-2r+1=0,$$
\commabsend
whose only solution is $r=1$, i.e., all servers crash in every game, which is obviously an equilibrium,
but not an interesting one. This proves that no equilibrium exists when there are three servers in the
game.

We expand this logic to make a general claim about the game.
\begin{theorem}\label{thm:crash}
For every $n\geq3$, the Player Failure Hotelling game $H_{\PF}(n,r,[0,1])$ has no Nash equilibrium in pure strategies.
\end{theorem}

\commful
\emph{Proof Sketch:} We observe that in this game interior servers must ``think'' like peripheral servers, i.e., each server $s_i$ would move away from a boundary and towards one of its neighbors.

This holds because each interior server $s_i$ has some probability of becoming a peripheral server, in which case $s_i$ would profit by increasing the hinterland. However, $s_i$ is more likely to become the peripheral server of the boundary closest to it (separated by less servers), and thus would do better in expectation by moving away from this boundary and towards the center.

It follows that all servers would converge towards the center, which clearly never results in an equilibrium.
\commfulend

\commabs
\begin{proof}
We have shown the claim holds for $n=3$ so suppose $n\geq4$.
Suppose there exists a Nash equilibrium in pure strategies for $H_{\PF}(n,r,[0,1])$, and let the three leftmost servers be $s_1$, $s_2$ and $s_3$, ordered from left to right (see Figure \ref{fig:Hgame-np}).
Let $s_1$'s hinterland be of length $\loc_1$,
let $s_2$ be paired with $s_1$ (otherwise $s_1$ would move closer to $s_2$)
and let $s_3$ be located at $\loc_3$.

\begin{figure}[ht]
\includegraphics[width=\textwidth]{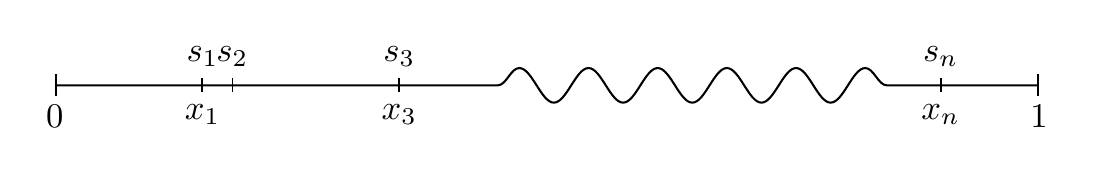}
\caption{The three leftmost servers in an n-server game.}
\label{fig:Hgame-np}
\end{figure}

Let us consider a move of $s_2$ within the interval between its present neighbors $s_1$ and $s_2$. In all fault scenarios where $s_2$ remains an interior server, such a move has no effect. In case $s_2$ either crashes or remains as the only non-faulty server, its location does not matter. But in cases where $s_2$ is a peripheral server, moving and increasing its hinterland would increase its payoff. However, $s_2$ is much more likely to become the left peripheral server than the right peripheral server, since the former requires only $s_1$ to crash while the latter requires $n-2$ servers to crash. For this reason, $s_2$ would profit by moving closer to $s_3$. A detailed formal analysis follows.

Suppose $s_2$ pairs with $s_3$ and consider the following cases:
\begin{enumerate}
\item \label{c1}
  \emph{$s_2$ becomes the left peripheral server and has a non-faulty neighbor $s_i$.}
  In this case pairing with $s_3$ increases $s_2$'s payoff from $ \loc_1 + (\loc_i-\loc_1)/2 $
  to $ \loc_1 + (\loc_3-\loc_1) + (\loc_i-\loc_3)/2 $.
  This occurs with probability $r \cdot (1-r)^2$.
\item \label{c2}
  \emph{$s_2$ becomes the right peripheral server and $s_1$ is non-faulty}.
  In this case pairing with $s_3$ decreases $s_2$'s payoff from $1-\loc_1$
  to $1-\loc_1-(\loc_3-\loc_1)/2$.
  This occurs with probability $r^{n-2} \cdot (1-r)^2$.
\item \label{c3}
  \emph{Otherwise,} pairing with $s_3$ does not change $s_2$'s payoff, because $s_2$ either crashed or remained the only non-faulty server or remained an interior server.
\end{enumerate}

In conclusion, when $s_2$ pairs with $s_3$, in case \ref{c1}
his payoff is increased by $(\loc_3-\loc_1)/2$ with probability $r \cdot (1-r)^2$,
and in case \ref{c2} his payoff is decreased by $(\loc_3-\loc_1)/2$ with probability $r^{n-2} \cdot (1-r)^2$. In cases \ref{c3}, $s_2$'s payoff is unchanged.
In total, $s_2$'s payoff increases as a result of this relocation by
$$
  r \cdot (1-r)^2 \cdot \frac {\loc_3-\loc_1}2 - r^{n-2} \cdot (1-r)^2 \cdot \frac {\loc_3-\loc_1}2    = (r - r^{n-2}) \cdot (1-r)^2 \cdot \frac {\loc_3-\loc_1}2~,
$$
which is strictly positive for $0<r<1$ and $n \geq 4$ as long as $\loc_3>\loc_1$.

Therefore, $s_2$ must be paired with $s_3$. However, the peripheral server $s_1$ must be paired with $s_2$ as well. But then $s_2$ would be paired on both sides and
as we have seen in the three server game this cannot be an equilibrium.
Namely, $s_2$ would pair with $s_1$ on the left or with $s_3$ on the right.
We thus reach a contradiction, proving that no equilibrium exists if there are more than
three servers in the game.
\end{proof}
\commabsend

\section{Conclusion} \label{sec:conclusion}

In this paper we considered two fault tolerant variants of the Hotelling game: link failures and player faults. On the one hand, we have shown that the game is resilient to link failures in some sense - each equilibrium is related to an equilibrium of the no-faults Hotelling game by rescaling the interval along with player positions. On the other hand, we have shown that the game is vulnerable to player failures. No equilibrium exists because players tend to converge towards the center.
\commabs
Table~\ref{tbl:comp} compares the numbers of Nash equilibria in each model

\begin{table}[h]
\caption{The number of Nash equilibria that exist in each variant of the game depending on the number of players.}\vspace*{1ex}
\begin{centering}
\begin{tabular}{|c|c|c|c|}
\hline
n & Server Crashes & Line Disconnect & No Faults\\
\hline
\hline
1 & $\infty$ & 1 & $\infty$\\
\hline
2 & 1 & 1 & 1\\
\hline
3 & 0 & 0 & 0\\
\hline
4 & 0 & 1 & 1\\
\hline
5 & 0 & 1 & 1\\
\hline
$\geq6$ & 0 & $\infty$ & $\infty$\\
\hline
\end{tabular}
\par\end{centering}
\label{tbl:comp}

\end{table}
\commabsend

There are many possible future directions for this research. A large number of variants of the Hotelling game have been studied and each would be interesting to consider in a faulty setting, such as: the Hotelling game on graphs, on the plane or over $\mathbb{R}^n$, the Hotelling game with sequential entry, and so on. Another interesting direction would be to try other fault models. Some examples are models where the number of faulty players is bounded, where faulty players remain in the game but act unexpectedly (or ``Byzantinely''), or where faults are injected adversarially rather than at random.

\commabs
\clearpage
\commabsend
{\small
\bibliographystyle{plain}
\bibliography{bib}
}


\commabs

\clearpage
\centerline{\Large \bf Appendix}

\appendix

\section{Nash Equilibrium of the 4-Server Game with Link Failures}
\label{app:nash4}

Consider the solution presented in Figure~\ref{fig:Hgame-4p-Disconnections}, and suppose that one of the servers, denoted $s^*$, moves to location $ y $ between $0$ and $x$. The identity of $s^*$ does not change his payoff since one of the other players will be located at $x$.

The following table compares the payoff of $s^*$ to the payoff of $s_1$ in the original state (wlog, because all servers have the same expected payoff).

\begin{center}
\begin{tabular}{| c | c | c | c |}
  \hline
  $f$ & $\Prb[f]$ &  $ s_1 $ &  $ s^* $ \\  \hline
  --- & $1-r$ & $x $ & $(x+y)/2$ \\
  $[0,y]$ & $ ry $ & $x - y/2 $ & $x/2$ \\
  $[y,x]$ & $ r(x-y) $ & $ (x-y)/2 $ & $ (x+y)/2 $ \\
  $[x,1]$ & $ r(1-x) $ & $x $ & $(x+y)/2$ \\
  \hline
\end{tabular}
\end{center}

\vspace{2ex}

We claim that $\Expct[p(s_1)] \geq \Expct[p(s^*)]$, and thus obtain the following inequality.

\begin{multline*}
(1-r) \cdot x + ry \cdot (x- \frac{y}{2}) + r(x-y) \cdot \frac{x-y}{2} + r(1-x) \cdot x
\geq \\ (1-r) \cdot \frac{x+y}{2} + ry \cdot \frac{x}{2} + r(x-y) \cdot \frac{x+y}{2} + r(1-x) \cdot \frac{x+y}{2} \text{.}
\end{multline*}

Shifting all terms to the left hand side and simplifying yields
\[
(1-r)\cdot(\frac{x-y}{2})+r \cdot \frac{x-y}{2} \cdot (1-y-x) \geq 0 \text{.}
\]

Recall that $ 0 < r < 1 $ and $ 0 \leq y < x < 1/2 $, therefore all terms
on the left hand side are non-negative and the inequality holds. Hence,
no server would gain from moving between $0$ and $x$.

It is left to show that no server would gain by moving between $ x $ and $ 1/2 $,
since the right half of the segment is symmetrical.

Suppose now server $s^*$ relocates at $ y $ between $ x $ and $ 1/2 $, we then obtain the following
table.

\begin{center}
\begin{tabular}{| c | c | c | c |}
  \hline
  $f$ & $\Prb[f]$ &  $ s_2 $ &  $ s^* $ \\  \hline
  --- & $ 1-r $ & $ 1/2 - x $ & $ 1/2 - x $ \\
  $[0,x]$ & $ rx $ & $ 1/2 - x $ & $ 1/2 - x $ \\
  $[x,y]$ & $ r(y-x) $ & $ (y-x)/2 $ & $ 1/2 - x $ \\
  $[y,1-x]$ & $ r(1-x-y) $ & $(1+y-3x)/2 $ & $ 1/2 - x $ \\
  $[1-x,1]$ & $ rx $ & $ 1/2 - x $ & $ 1/2 - x $ \\
  \hline
\end{tabular}
\end{center}

\vspace{2ex}

We conclude that any server moving between $x$ and $1/2$ would have an expected payoff of
$ 1/2 - x $, which does not improve his payoff (rather, his payoff remains unchanged).

In summary, when the servers $ s_1 $, $ s_2 $, $ s_3 $ and $ s_4 $ are located at
$\loc_1=\loc_2=(2-\sqrt{4-r})/r$ and $\loc_3=\loc_4=1-(2-\sqrt{4-r})/r$ respectively (see Figure~\ref{fig:Hgame-4p-Disconnections}), the system is at a Nash equilibrium.
No server may improve his expected payoff by changing his location (assuming the other servers stay in the same place).

\commabsend

\end{document}